# Multifilamentary character of anticorrelated capacitive and resistive switching in memristive structures based on $(CoFeB)_x(LiNbO_3)_{100-x}$ nanocomposite

M.N. Martyshov[1,a)], A.V. Emelyanov[2,3], V.A. Demin[2,3], K.E. Nikiruy[2], A.A. Minnekhanov[2], S.N. Nikolaev[2], A.N. Taldenkov[2], A.V. Ovcharov[2], M.Yu. Presnyakov[2], A.V. Sitnikov[2,4], A.L. Vasiliev[2,3], P.A. Forsh[2,3], A.B. Granovsky[1], P.K. Kashkarov[1,2,3], M.V. Kovalchuk[1,2,3], V.V. Rylkov[2,5,6,b)]

[1]Faculty of Physics, Lomonosov Moscow State University, 119991 Moscow, Russia

[2]National Research Center "Kurchatov Institute", 123182 Moscow, Russia

[3]Moscow Institute of Physics and Technology (State University), 141700 Dolgoprudny, Moscow Region, Russia

[4]Voronezh State Technical University, 394026 Voronezh, Russia

[5]Kotelnikov Institute of Radio Engineering and Electronics RAS, 141190 Fryazino, Moscow Region, Russia

[6]Institute of Applied and Theoretical Electrodynamics RAS, 127412 Moscow, Russia

a) mmartyshov@mail.ru; b) vvrylkov@mail.ru

**Abstract**

Resistive and capacitive switching in capacitor metal/nanocomposite/metal (M/NC/M) structures based on $(CoFeB)_x(LiNbO_3)_{100-x}$ NC fabricated by ion-beam sputtering with metal content $x \approx 8$–$20$ at. % is studied. The peculiarity of the structure synthesis was the use of increased oxygen content ($\approx 2 \cdot 10^{-5}$ Torr) at the initial stage of the NC growth. The NC films, along with metal nanogranules of 3-6 nm in size, contained a large number of dispersed Co (Fe) atoms (up to $\sim 10^{22}$ cm$^{-3}$). Measurements were performed both in DC and AC (frequency range 5–13 MHz) regimes. When switching structures from high-resistance ($R_{OFF}$) to low-resistance ($R_{ON}$) state, the effect of a strong increase in their capacity was found, which reaches 8 times at $x \approx 15$ at. % and the resistance ratio $R_{OFF}/R_{ON} \approx 40$. The effect is explained by the synergetic combination of the multifilamentary character of resistive switching (RS) and structural features of the samples associated, in particular, with the formation of high-resistance and strongly polarizable LiNbO$_3$ layer near the bottom electrode of the structures. The proposed model is confirmed by investigations of RS of two-layer nanoscale M/NC/LiNbO$_3$/M structures as well as by studies of the magnetization of M/NC/M structures in the pristine state and after RS.

# I. INTRODUCTION

Structures of the type of metal/oxide/metal (MOM), demonstrating the effects of reversible resistive switching (RS), are very promising for creating multilevel memories and memristor arrays to emulate synapses in the development of neuromorphic (bio-inspired) computational systems to solve the so-called anthropomorphic problems: image and natural language recognition, decision-making, generalization, prediction, etc. [1–11]. Such systems possess low power consumption and can essentially surpass modern computer systems based on von Neumann architecture by efficiency of information processing (speed and reliability) at performance of the cognitive functions listed above.

Over the last decade, a sufficiently large number of different MOM memristive structures were discovered and studied. They can be divided into several basic types depending on the RS mechanism. First of all, it should be noted that the RS effects in the most studied MOM structures are associated with the processes of electromigration of oxygen vacancies in the oxide layer ($TiO_x$, $TaO_x$, $HfO_x$, etc.), or metal cations such as Cu, Ag into the dielectric (e.g., $SiO_2$) from the active electrode of the MOM structure [1, 7–15]. In these cases, RS of the structure to the low resistive state (LRS) is caused by the formation of (i) conductive filaments (during the electro-migration of oxygen vacancies); or (ii) conductive metal bridges (during the migration of metal cations) in the oxide matrix, closing the contacts of the MOM structure; or (iii) suppressing (bypassing) the Schottky barrier at one of the structure electrodes. The latter is usually formed using semiconductor oxides with a band gap of $\leq 3$ eV. When the voltage polarity is reversed, the filament (bridge) is destroyed and a reversible RS to the high resistance state (HRS) is observed in the structure. For stable operation of anionic (with a change of valence) or cationic memristive MOM structures the electroforming process at high voltages is usually required for the first filaments (bridges) formation. However, other RS mechanisms, which do not require electroforming, are possible, such as the recharge of localized electronic states in the Schottky barrier region and/or in the oxide volume (see [16, 17] and Refs. therein), electric repolarization of ferroelectric oxide [18–20], redox reactions in organic materials [21], electron drag effect of metal atoms [22] and, finally, temperature-induced metal-insulator transition in Mott materials ($NbO_2$, $VO_2$, $V_2O_3$), which, although was discovered relatively long ago, is still the subject of discussion and studies [8, 23].

Despite the considerable amount of accumulated experimental results, there is still no microscopic theory of reversible RS. In particular, this is due to the difficulties in describing the interrelated non-equilibrium processes of thermal, electronic, and ion transport in nanometer scale, as well as due to the participation of several RS channels associated, for example, with the simultaneous manifestation of cationic and anionic transport [14] or with the synergetic contributions to RS of ion transport and electric polarization of ferroelectrics [20].

In recent years considerable attention has been paid to the study of phenomena that can accompany the effects of RS in the MOM structures and clearly reflect their features. Such phenomena include: integer (multiple $G_0$) and/or half-integer ($G_0/2$) quantization of the structure conductance at room temperature ($G_0 = 2e^2/h$, where $e$ is the electron charge, and $h$ is Planck's constant) [24-26], and also the change (switching) of its capacity $C$ during the RS [16, 17, 27–32].

The currently known capacitive switching (CS) mechanisms can be divided into two types. The first of them is related to the recharging of traps and/or movement of oxygen vacancies in the Schottky barrier region, which change its size and, accordingly, the capacity of MOM structure [16, 17, 27]. The other is due to the formation (destruction) of the filament percolation network during the movement of vacancies, which changes the effective permittivity and resistance of the MOM structure [28–32]. In the first case, the capacitance change during RS can be significant and exceed 10 times [16, 27]. The work [16] reports as high as a 100-fold change in the capacity of the Au/DyMnO$_3$/Nb:SrTiO$_3$/Au structures under RS. However, such a strong CS, apparently, is entirely due to the introduction into the MOM active region of the DyMnO$_3$/Nb:SrTiO$_3$ heterojunction structure, which has a large capacity. The change in the Schottky barrier capacity at the Nb:SrTiO$_3$/Au interface at RS does not exceed 3 times (see inset to Fig. 2c in [16]). Under the second mechanism, $C$ variation typically does not exceed 3–4 times, and the effects of CS and RS in this case can be both correlated (signs of change of $C$ and $R$ are the same) [29–31], and anticorrelated [28, 32] (in [31] at large $R \geq 10^4$ Ω growth of $C$ is followed by a fall).

Recently, we have revealed significant effects of RS in metal/nanocomposite/metal (M/NC/M) structures based on a (Co$_{40}$Fe$_{40}$B$_{20}$)$_x$(LiNbO$_3$)$_{100-x}$ NC consisting of 2–3 nm CoFe nanogranules in the amorphous nonstoichiometric LiNbO$_{3-y}$ matrix with a high content of dispersed Fe$^{2+}$ and Co$^{2+}$ magnetic ions, up to $n_i \sim (2-3) \cdot 10^{22}$ cm$^{-3}$ at $x \approx 33-47$ at.% [33, 34]. The strongest RS in M/NC/M structures is observed at the metal phase content $x \sim 10$ at. % with the ratio of resistance in high-resistance state to that in low-resistance one $R_{OFF}/R_{ON}$ up to $\sim 10^2$. Under these conditions the number of stable RS cycles (endurance) can exceed $10^6$ and retention time of resistance state is more than $10^4$ s [35-37]. But the most interesting property of these memristive structures is the analog nature of their switching or their high degree of plasticity (it is possible to set them for more than 256 resistance states) [35]. The latter property allowed us to implement various possibilities for changing the memristor conductance (its weight) according to bio-inspired rules like STDP (spike-timing-dependent plasticity) and demonstrate adaptive (self-learning) properties of NC memristors when using them as synapses for implementing spiking neural networks (SNNs) [35-37]. Note that the high level of plasticity of memristors reduces the requirements to their endurance ($\geq 10^5$) [38] and the window of resistance changes ($R_{OFF}/R_{ON} \geq 10$) [5, 6] for neuromorphic applications.

Another interesting feature of M/NC/M memristors is the possibility of using sufficiently thick layers of NC ($d \approx 1-3$ μm) when fabricating them, while maintaining relatively small switching voltages from the HRS to LRS and vice versa: $U_{LRS}$, $U_{HRS} \approx 3\text{-}5$ V [33, 35, 36, 39], respectively. This peculiarity makes it possible to produce memristive arrays of crossbar structures based on simple technologies using electric bus with a typical thickness of ~0.1 μm [40], as well as, apparently, to avoid the problems of structure degradation associated with the influence of moisture and interface gas bubbles on the RS [41, 42].

Recently, we also found that RS is strongly suppressed in M/NC/M structures ($x \approx 15$ at.%, $d \approx 1.4$ μm) under conditions when the concentration of dispersed magnetic ions in the insulating matrix falls to a low level of $n_i \sim 10^{20}$ cm$^{-3}$ [43]. Under these conditions, the RS are unstable, and the resistance ratio $R_{OFF}/R_{ON}$ does not exceed 5.

The RS mechanism in M/NC/M structures remained unclear in [33, 35, 36, 43]. However, based on the obtained results, it can be assumed that the observed RS effects are determined by the local injection (or extraction) of oxygen vacancies (depending on the voltage sign) into a strongly oxidized NC layer near the electrode of the structure controlling its resistance. Regions of local injection are defined by the positions of percolation chains of granules surrounded by vacancies and/or magnetic ions [33] ("metallized" granular chains), which do not change in the process of RS, thus ensuring their high stability. On the other hand, the high level of plasticity of NC memristive structures can be associated with the multifilamentary character of RS.

To prove the last assumption, we performed capacitive and magnetic measurements along with RS ones on M/NC/M structures specially fabricated for these purposes. We found out that changes in the capacity of M/NC/M structures can reach 8 times during the RS process. The observed effect of CS is explained by the formation of numerous nanocapacitors near the bottom electrode of the M/NC/M structure, with upper plates determined by the position of metallized granular chains (MGCs), and by a layer of amorphous/nanocrystalline LiNbO$_3$, whose permittivity reaches $\varepsilon_d \sim (10^2-10^4)$ [44]. According to magnetic measurements of M/NC/M structures before and after RS, the volume fraction of the resulting MGCs at the stage of "soft electroforming" is about $K_v \sim 1\text{-}10\%$, which also indicates the multi-channel nature of RS with a surface density of filaments of $\sim 10^{10} - 10^{11}$ cm$^{-2}$.

## II. SAMPLES AND EXPERIMENTAL DETAILS

M/NC/M structures based on the (Co$_{40}$Fe$_{40}$B$_{20}$)$_x$(LiNbO$_3$)$_{100-x}$ NC are synthesized by ion-beam sputtering using a composite target consisting of cast alloy Co$_{40}$Fe$_{40}$B$_{20}$ (CoFeB) plate and 14–15 strips of ferroelectric LiNbO$_3$ (see details in [33]). An elongated rectangular target with a non-uniform arrangement of LiNbO$_3$ strips was used, allowing the formation of NC with different concentration of the metal phase in the range of $x$ = 6–43 at. % (with the accuracy of $\delta x$ ~ 0.5-0.8 at. %) in a single cycle of synthesis. NC was deposited in the argon atmosphere ($P_{Ar} \approx 8\cdot10^{-4}$ Torr) through a shadow mask with the hole diameter of 5 mm at room temperature on the glassceramic substrates covered with a metal film. The latter served as the bottom electrode (BE) of the M/NC/M structures. Note that the ion-beam sputtering of NC in comparison with the magnetron sputtering does not require RF and special cooling of substrates (the deposition process takes place outside of the hot plasma).

For capacitive measurements, we used M/NC/M structures (CM M/NC/M structures) with a developed oxide interlayer between NC and the bottom electrode. For this purpose at the initial stage, the NC deposition was performed in the mode of a given oxygen flow at its increased partial pressure $P_{O2} \approx 2.1\cdot10^{-5}$ Torr for 8 min, after which the flow of $O_2$ was decreased. Further deposition was carried out for 120 min at an average pressure of $P_{O2} \approx 1.4\cdot10^{-5}$ Torr ($P_{O2}$ values are given for the chamber vacuum limit $P \approx 6\cdot10^{-6}$ Torr). The thickness of the NC layer in the structures was $d \approx 1.5$ μm. The electrodes of the structures were made of three-layer Cr/Cu/Cr metal film with a thickness of $\approx 1$ μm.

Note that usually the first stage of NC synthesis takes no more than 4 min at high $O_2$ partial pressure of about $2\cdot10^{-5}$ Torr; in this case, the typical switching voltages are: $U_{LRS}$, $U_{HRS} \approx$ 3-5 V [33, 35, 36, 39]. This approach was used to fabricate M/NC/M structures ($d \approx 1.2$ μm) for magnetic measurements (MM M/NC/M structures) with non-magnetic Cu contacts (M is Cu).

Additionally, layered capacitor structures based on films of NC $(CoFeB)_x(LiNbO_3)_{100-x}$ and $LiNbO_3$ (M/NC/LiNbO$_3$/M structures) with thicknesses of 10 and 40 nm, respectively, were fabricated. LiNbO$_3$ was formed at the BE of the structure; Cr/Cu/Cr films were used as electrodes.

The top electrodes (TEs) in all the cases described above were deposited through the shadow mask with periodically located holes of size $S$ = 0.5×0.2 mm$^2$, the surface fraction of which was $K_s \approx 0.25$.

It should be noted that the above formula for the NC was used by us to find $x$ according to the data of energy-dispersive X-ray (EDX) microanalysis [33]. In fact, a significant part of boron appears in the insulating matrix, outside the granules, in the process of synthesis of the NC [33, 45]. At the same time, it is impossible to determine which part of boron remains in the granules by the existing EDX spectroscopy methods. Therefore, following [33, 34], we will use the NC formula notation, reflecting the composition of the target, denoting it for short as $(CoFeB)_x(LiNbO_3)_{100-x}$.

The microstructure of the NC films were studied by high-resolution transmission and scanning transmission electron microscopy (TEM and STEM, respectively) using a TITAN 80-300 TEM/STEM instrument (ThermoFisher Scientific, USA) operating at an accelerating voltage of $U$ =

300 kV, equipped with a Cs-probe corrector, high-angle annular dark-field detector (HAADF) (Fischione, US) and EDX microanalysis spectrometer (EDAX, USA). For the image processing Digital Micrograph (Gatan, USA) software and TIA (ThermoFisher Scientific, USA) were used. Details of preparation of samples for structural studies with atomic resolution are described in [45].

The capacitance of the structures was measured with the help of the HP 4192A impedance analyzer in the frequency range $f$ from 5 Hz to 13 MHz at amplitude of an alternating signal of 50 mV. Investigations of the electrophysical properties of the M/NC/M structures at DC, including measurements of their I–V characteristics, were carried out with the help of a four-channel source measure unit PXIe-4140 (National Instruments) using the analytical probe station PM5 (Cascade Microtech). In the study of the I–V curves, current $I$ was measured with the grounded bottom electrode of the structure and with the alternating voltage $U$ applied to the top electrode according to the linear law in the sequence from $0 \rightarrow + U_0 \rightarrow - U_0 \rightarrow 0$ V with the step of 0.1 V and the amplitude of saw-tooth sweep up to $U_0 = 15$ V. In addition, we studied the temperature dependence of both the I–V characteristics of these structures and their conductance in the relatively weak fields ($\leq 10^3$ V/cm) in the temperature range $T = 10$–300 K using an evacuated insert, immersed in a liquid-helium Dewar flask.

The magnetic properties of the Cu/(CoFeB)$_x$(LiNbO$_3$)$_{100-x}$/Cu structures were studied with a Quantum Design MPMS-XL7 SQUID magnetometer at $T = 2$–300 K and a magnetic field up to 7 T oriented in the sample plane.

## III. RESULTS AND DISCUSSION

### A. Conductance of the CM M/NC/M sandwiches vs metal content and their resistive switching

Fig. 1a shows the dependence of the conductance of the CM M/NC/M structures, $G(x) = I/U$, on the metal content measured at DC at $U = 0.3$ V, after action to the saw-tooth voltage with the amplitude up to $U_0 = 3$–5 V, which is insufficient for RS of the structures to the LRS ($U_{LRS} \approx 10$ V; see the inset to Fig. 1a). At the same time, however, a significant increase in the stability and reproducibility of the measured value $G$ was provided. The dependence of $G(x)$ is typical for percolation granular systems [46]: below some threshold value $x < x_p$ (percolation threshold; in our case $x_p \approx 17$ at. %) the function $G(x)$ is exponential, and above it $G(x)$ practically does not depend on $x$.

According to our recent studies [33, 35, 39], the strongest and most stable RS in the M/NC/M structures is observed at some optimal content of metal $x_{opt}$ below the percolation threshold. The specific value of $x_{opt}$ may depend on the size of the granules and the anisotropy of their shape (elongation along the NC growth axis). In the synthesized CM M/NC/M structures, the maximum $R_{OFF}/R_{ON}$ ~40 ratio is observed at the $x_{opt}$ value of about 15 at. % (see I–V curves in the inset to Fig. 1a), which is close to the $x_{opt}$ value of ~10 at. % obtained for M/NC/M structures with $R_{OFF}/R_{ON}$ ~ 50 and strongly elongated granules (up to 10 nm with lateral sizes of 2-4 nm) [33, 39]. However, the switching voltage $U_{LRS} \approx 10$ V was almost 2-3 times higher than in the case of the structures studied in [33, 35, 36, 39].

Fig. 1b shows for comparison the I–V curves for the sufficiently thick M/NC/M structure ($d \approx 2.5$ μm) synthesized as in [33] using an excess atmosphere (bad vacuum) at the initial stage of NC growth. From the presented data, it is clear that in this case the M/NC/M structure has a noticeably lower switching voltage $U_{LRS} \approx 4$ V, that indicates a crucial role in RS the oxide interlayer formed at the bottom electrode (this is true for structures with a high concentration of dispersed magnetic atoms $n_i \sim 10^{22}$ cm$^{-3}$ [43]). For structures with $U_{LRS} \approx 3$-4 V, the endurance of RS exceeds $10^5$-$10^6$ (see the inset to Fig. 1b and data of [35, 36, 39]; the study of endurance is limited in our case by the possibilities of the measuring system used, i.e., the difficulties of measurements at pulse durations less than 10 ms). The switching voltage $U_{LRS}$ in such structures depends on the metal content and can fall up to several times in the immediate vicinity of the percolation threshold (see Fig. 1b in [35]). When comparing the curves in the Figs 1a and 1b, one can also see that in both cases there is a maximum in the region of negative voltages $U = 2$-3 V. Such maximum is due to the filament destruction and is observed in many types of MOM structures with bipolar RS based on the valence change mechanism (see, for example, [14, 47-49]). On the other hand, some current burst observed in the I-V characteristics for CM M/NC/M structure at positive voltage RS to the LRS (see Fig. 1a) is probably caused by the presence of bottlenecks in the formed MGCs, which are destroyed when the current passes, for example, due to local thermally-accelerated diffusion and oxidation processes [50]. With temperature lowering, the current in such MGCs decreases and, at the same time, the thermal effects are suppressed, which leads to the monotonous behavior of RS (see Fig. 9 in Sec. G).

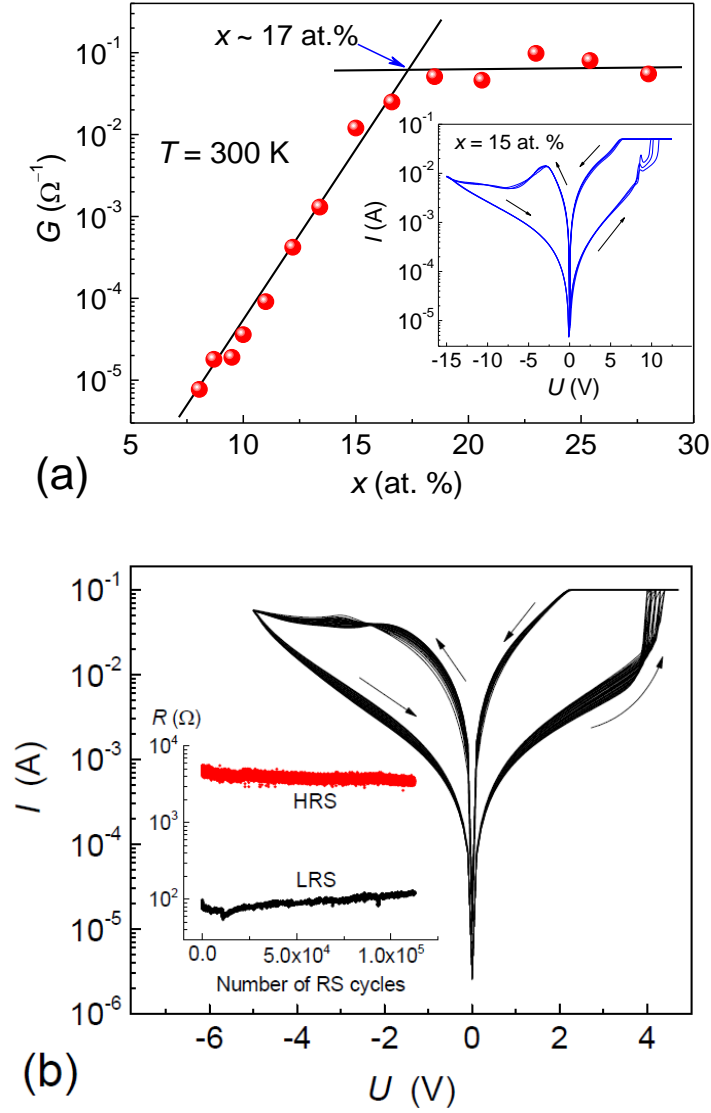

**Fig. 1.** (a) The dependence of conductance $G(x) = I/U$ of the CM M/NC/M structures ($d \approx 1.5$ μm) on the metal content $x$, measured at DC at $U = 0.3$ V. The inset shows the I-V curves for the CM M/NC/M structure with optimal content of $x_{opt} \approx 15$ at. %. (b) The I-V characteristics of the M/NC/M structures with the NC thickness of $d \approx 2.5$ μm and metal content of $x \approx 8$ at. % obtained by 30 cyclic measurements. The inset demonstrates the endurance to RS of this structure, i.e. the dependence of resistance after set and reset pulses on the switching cycle number. The arrows on the insets show the voltage scanning direction.

Note that the studies of cyclic RS of the structures were performed at rather large compliance currents $I_c$ = 50-100 mA (Fig. 1); the compliance current reached the same value in our recent studies [33, 35-37, 39]. If we assume that this current in the M/NC/M structure flows through a single MGC, the lateral size of which slightly exceeds the diameter of the granules and is ~10 nm (as for the filaments of most memristors [12-14, 31]), the current density in such MGC would reach approximately $10^{11}$ A/cm$^2$. This is a huge and unrealistic value compared to the maximum current density ~$(10^6$-$10^7)$ A/cm$^2$ that good metal microconductors can withstand without destruction due to the electromigration effects [22, 51]. From this simple consideration, it follows that at RS of the M/NC/M structure to the LRS, its conductance is determined by numerous MGCs (>$10^5$).

## B. Microstructure of the CM M/NC/M sandwiches with resistive switching effect

The bright field STEM images of the CM M/NC/M sandwich with optimal value $x_{opt} \approx 15$ ат.% at different magnifications are shown in Fig. 2a. The CoFe nanogranules embedded in LiNbO$_3$ layer are sandwiched between Cr/Cu/Cr layers. Close inspection of the interface between the upper Cr layer of the bottom electrode (BE) and LiNbO$_3$ with CoFe nanogranules (marked as CoFe-LiNbO$_3$) revealed an amorphous layer with the thickness of $d_{ox} \approx 15$ nm. The enlarged HAADF STEM image of this area is presented in Fig. 2b. The EDX microanalysis unambiguously demonstrated 25 at.% of Nb and 75 at.% of O stoichiometry in this amorphous layer so the layer exhibits dark contrast in Fig.2b. Of course, Li cannot be detected by EDX microanalysis. The results of elemental scan along the white line (from top to bottom) are shown in Fig 2c. The interface between the bottom electrode Cr layer and LiNbO$_3$ is relatively abrupt, the interdiffusion was not observed.

Previously [33], we found that in HAADF STEM images, the CoFe granules have a noticeably brighter contrast than the LiNbO$_3$ matrix. To prove that the areas with bright contrast correspond to CoFe granules we performed the detailed EDX mapping for Fe and Co elemental distribution [34]. There is the unambiguous match of Fe and Co distribution and these areas correspond to the bright areas in HAADF STEM image (see Fig. 2 in [34]). Figs 2d and 2e show HAADF STEM images of bottom part (e) of the composite and its upper one (d) located at the distance of about 100 nm from the BE. The corresponding histograms of the granule size distribution are shown in Figs. 2g and 2f. The bottom part of composite CoFe-LiNbO$_3$ layer consists of CoFe granules, which are approximately 6 nm in size and are significantly larger than the ones in the upper part of the NC layer. That part of the layer turns out to be the NC layer with a homogeneous distribution of CoFe nanogranules with the size of $a_g \approx 3$ nm (Fig. 2f). The high resolution (HR) TEM image (Fig. 2h) demonstrated that the LiNbO$_3$ matrix is amorphous and the CoFe nanogranules are crystalline. The 2D Fast Fourier Transform (FFT) spectrum corresponding to the CoFe granule marked in Fig.2h by red arrow and the HRTEM image of its crystal lattice in [001] zone axis are shown in the Figs. 2i and 2k, respectively. The last image together with the FFT spectra indicated that the CoFe nanogranules have a bcc crystal structure with lattice parameter $a = 0.29$ nm. The {110} crystal planes of bcc lattice are clearly visible with interplanar distance of ~0.21 nm.

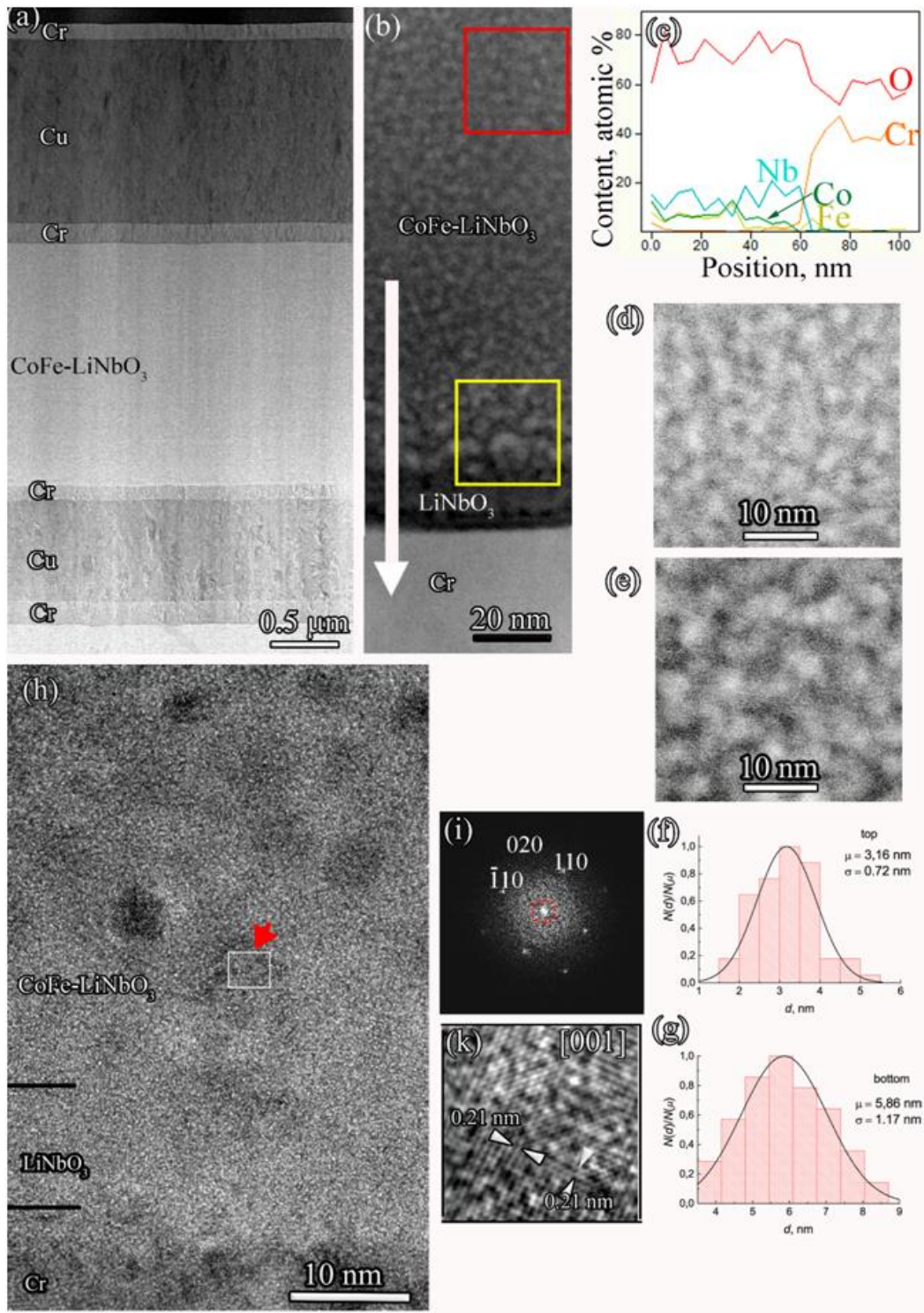

Fig.2. The cross-section image, results of EDX analysis and size distribution of granules of CM M/NC/M structure with optimal value $x_{opt} \approx 15$ at. %. (a) Bright field STEM image of the sample. (b) Enlarged HAADF STEM image of the area near bottom electrode. The white line demonstrates the area of EDX element analysis, shown in (c). (d), (e) HAADF STEM images of the areas marked in (b) with red and yellow squares, respectively. (f), (g) Histogram of the transverse-size distribution of granules and Gaussian approximation of this dependence (solid curve) for the images shown in (d) and (e), respectively. (h) HRTEM image of the interface near the bottom electrode. (i) The 2D FFT spectrum corresponding to the CoFe granule marked in (h) by the red arrow. (k) The image of the granule crystal lattice in [001] zone axis.

## C. Behavior of the CM M/NC/M structures capacity in different resistive states

Fig. 3 shows a typical impedance hodograph (Cole–Cole plot), i.e. the dependence of the imaginary part of impedance $Z''$ on the real one $Z'$, which was obtained for CM M/(CoFeB)$_x$(LiNbO$_3$)$_{100-x}$/M structure with $x \approx 15$ at. % in the HRS. The inset to Fig. 3 also shows the dependence of $Z''(Z')$, obtained for the same structure in the LRS. Similar dependencies were observed for all the CM M/NC/M structures with different content of metal phase. Note, that the hodographs obtained for the HRS and LRS look like semi-circles.

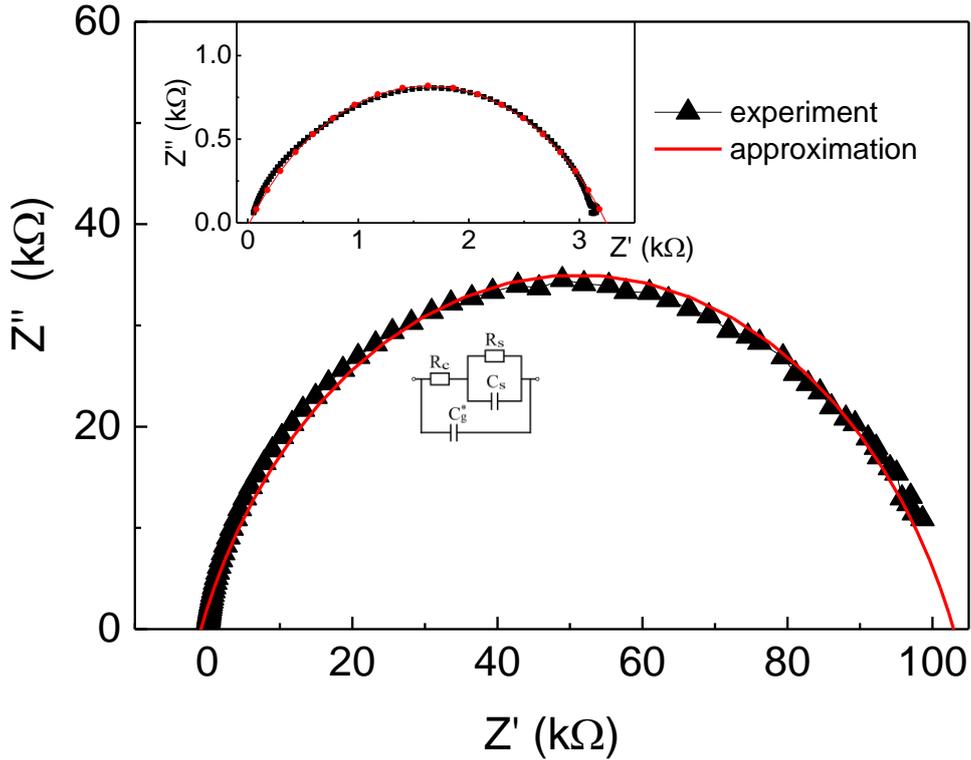

**Fig. 3.** The dependences of the imaginary part of the impedance on the real one for the CM M/NC/M structure with $x \approx 15$ at. % in the HRS and LRS (the top inset). The bottom inset shows the equivalent circuit of the CM M/NC/M structure (explanations are given in the text).

According to the qualitative model of RS proposed in [33], to describe these dependencies it would be necessary to use an equivalent electrical scheme (see the bottom inset in Fig. 3), consisting of a parallel $R_sC_s$ circuit and a resistance $R_c$ connected to it in series, which are shunted by a parallel connected geometric capacitance $C_g^*$ of the M/NC/M structure, modified because of the presence of metallic nanoparticles in the dielectric. The resistance $R_s$ and the capacitance $C_s$ are naturally related to the presence of a high-resistance layer near the BE of the structure, which determines its switching from one resistive state to another. The resistance $R_c$, in turn, is related to the resistance of the MGCs, providing contact of the $R_sC_s$ circuit with the top electrode of the structure. It is obvious that the resistance $R_c$ is much lower than the resistance of the switching layer $R_s$ in the HRS of the memristor.

It is known that an equivalent circuit, similar to that described above, should in general lead to two semi-circles on the hodograph [52]. However, this is not observed in our experiments because the time constant $R_c C_g^* \leq 10^{-8}$ s (see below) is rather small and lies far beyond the range of relaxation times ($1/f \approx 10^{-7}$ s). Therefore, the hodographs shown in Fig. 3 are described quite satisfactorily by the separate $R_s C_s$ circuits (corresponding curves are shown in solid lines in Fig. 3).

It should be noted that the centers of the semi-circles of impedance hodographs do not lie on the axis of the real Re(Z) impedance, but are below it. This fact may be related to the presence of some distributed elements in the nanocomposite/high impedance layer/electrode system. This leads to the conclusion that the relaxation time $\tau$ is not a constant value, but is continuously or discretely distributed with respect to its mean value [52].

In this situation, we should expect a fairly strong dependence of the measured capacitance on the frequency $C(f)$. In Fig. 4a the dependencies of $C(f)$ in the HRS and LRS of the structure on the alternating signal frequency for a sample with $x \approx 15$ at. % are shown.

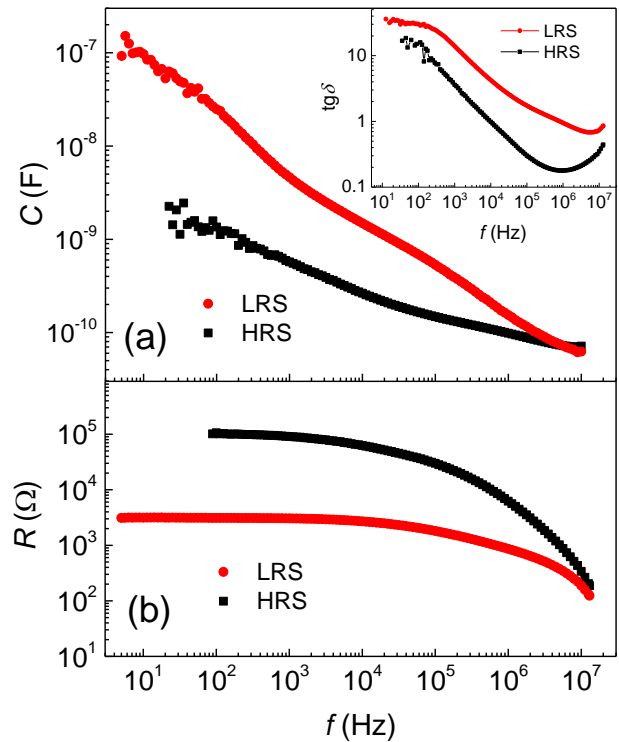

**Fig. 4.** The frequency dependences of capacitance (a) and resistance (b) in the HRS and LRS for the M/(CoFeB)$_x$(LiNbO$_3$)$_{100-x}$/M structure with $x \approx 15$ at. %. The inset shows the frequency dependences of loss tangent in the HRS and LRS for this structure.

As can be seen from the figure, the capacitance of the structure decreases in both cases with increasing the frequency. Meanwhile at high frequencies ($f \approx 10^7$ Hz) the capacitances in the HRS and LRS practically coincide. Similar dependencies of $C(f)$ were obtained for the other samples studied. It is necessary to note that the resistances in the HRS and LRS at the high frequency ($R_{HF}$) of $f \approx 10^7$ Hz also coincide (see Fig. 4b). And, according to the equivalent scheme (Fig. 3), $R_c \leq R_{HF} \sim 100$ Ω. In addition, note that the structure resistance, when below the percolation threshold, is almost frequency-independent up to $\sim 10^4$ Hz in the HRS and up to $\sim 10^5$ Hz in the LRS. Such behavior is typical for granulated NC films, including those with hopping conductivity, in which the real part of the conductance practically does not depend on the frequency up to $f \sim 10^4$ Hz (see, for example, [53]).

The tangent tg$\delta$ of dielectric losses is presented in the inset to Fig. 4a. In both resistive states, the tg$\delta$ value decreases monotonously with increasing frequency. However, at high frequencies, there is a clear tendency for the dissipation factor to increase, especially in the HRS. It is known that the drop in tg$\delta$ with the frequency is typical for a parallel $RC$ circuit with a weak dependence of resistance and capacitance versus frequency (tg$\delta = 1/\omega \cdot R \cdot C$, $\omega = 2\pi f$), whereas for a series $RC$ circuit under these conditions, an increase in the tg$\delta$ with $f$ should be observed (tg$\delta = \omega \cdot R \cdot C$) [52]. From this one can conclude that at $f$ less than $10^6$ Hz, the dissipation factor is mainly determined by the high-resistance oxide interlayer formed at the BE of the structure ($R_s C_s$ circuit in the inset to Fig. 3), which decreases with increasing frequency. However, at $f$ above $10^6$ Hz, losses probably begin to manifest in the MGCs ($R_c$ resistance in the inset to Fig. 3) and the tg$\delta$ tends to grow.

A significant decrease in capacitance with frequency growth (Fig. 4a) can be attributed both to a decrease in the permittivity of the amorphous LiNbO$_3$ matrix with an increase in the frequency of the alternating signal [44] and to the shunting effect of the effective geometric capacitance $C_g^*$ of the M/NC/M structure, which can be significantly increased due to the presence of metallic nanoparticles in the dielectric matrix. In the Maxwell-Garnett approximation, which is valid for a small volume fraction of spherical nanoparticles $x_v \ll 1$, the complex effective permittivity of the composite medium is determined by the expression [54]:

$$\varepsilon_{MG} = \varepsilon_d \frac{\varepsilon_m + 2\varepsilon_d + 2x_v(\varepsilon_m - \varepsilon_d)}{\varepsilon_m + 2\varepsilon_d - x_v(\varepsilon_m - \varepsilon_d)}, \qquad (1)$$

where $\varepsilon_d$ and $\varepsilon_m$ are permittivity of the dielectric and metal, respectively. Considering, as usual, that for metal the imaginary part of the permittivity Im($\varepsilon_m$) is much larger than its real part Re($\varepsilon_m$) and also much larger than the permittivity of the dielectric $\varepsilon_d$, which is obviously fulfilled up to high frequencies, we obtain that the real part of the effective permittivity of the composite medium is equal to:

$$\varepsilon_{eff} = \text{Re}\,\varepsilon_{MG} \approx \varepsilon_d(1+3x_v)\,. \tag{2}$$

The equivalent circuit in Fig. 3 shows that when the capacitance values in the HRS and LRS are close, the capacitance measured is defined as:

$$C \approx C_g^* \approx C_g(1+3x_v) = \frac{\varepsilon_d S}{4\pi d}(1+3x_v)\,, \tag{3}$$

where $C_g$ is the geometric capacitance of the structure in the absence of metallic nanoparticles in the dielectric. At $x \approx 15$ at. % the volume fraction of metal $x_v \approx 8$ vol. % for the $(CoFeB)_x(LiNbO_3)_{100-x}$ NC. Taking this into account, substituting the nominal value of $\varepsilon_d \approx 50$ [44] in (3), we obtain $C \approx 40$ pF, which is close to the experimentally measured value of $C \approx 60$ pF (Fig. 4a). It should be noted here that the elongation and/or short-circuit of the granules along the growth axis of the NC can strongly increase their polarizability and the total capacity of the structure. In a more general case, for ellipsoidal granules: $\varepsilon_{eff} \approx \varepsilon_d(1+x_v/L)$, where $L$ is a form-factor less than 1/3 for oblong ellipsoid ($L = 1/3$ for the sphere). The experimentally found value of $C = C_g^* \approx 60$ pF corresponds to the form-factor $L \sim 0.1$, inherent in the granules in the shape of prolate ellipsoid of revolution with the long to short axis ratio of $\sim 3$. Note that in our case $R_{HF} C_g^* \sim 10^{-8}$ s, which is much less than $1/f$.

The switching of the M/NC/M structure to the LRS should be accompanied by a reduction in the thickness of the high-resistance layer, which in turn should lead to an increase in capacitance. This effect, obviously, should be the stronger the greater the $R_{off}/R_{on}$ ratio, which is quite clearly manifested in our case in the study of correlation between the relative change in resistance and capacitance in the structures with different metal content (Fig. 5). The maximum relative capacitance change of up to 8 times, as well as the maximum relative resistance change (about 40 times), is observed below the percolation threshold for the structures with $x \approx 14$–16 at. %.

A fairly significant change in $C$ during RS with formation of metal filaments was observed in [31, 32], where capacity variations reached 3–4 times. Apparently, the multifilament mechanism of RS close to our case was realized in [31]. At the same time, in [31] the capacitance at first slightly increased with decreasing resistance, and then strongly (4 times) fell, which was associated with the formation of solid (without ruptures) metal Ni bridges. In our case, when the resistance decreases, the capacitance experiences only growth.

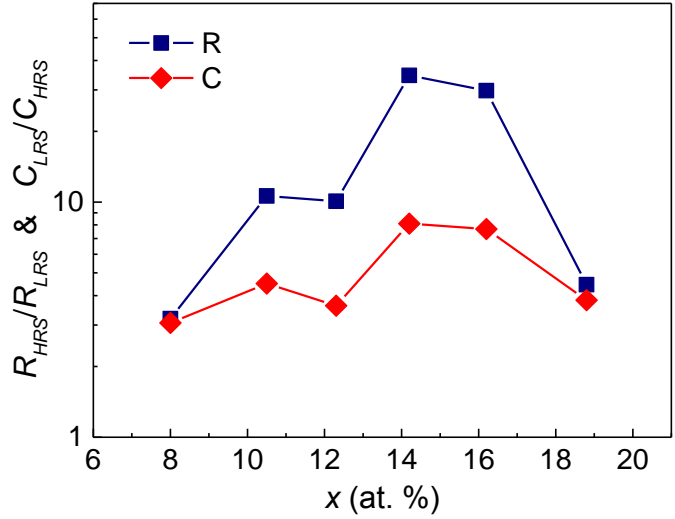

**Fig. 5.** Dependence of the relative change in resistance and capacity of the CM M/NC/M structure during its RS between the HRS and LRS on the content of metallic phase at a frequency of 1 kHz.

### D. Qualitative model of resistive/capacitive switching

Let one consider a possible multifilament mechanism of RS in our case, taking into account: (i) the high content of dispersed Co and Fe atoms in the insulating $LiNbO_{3-y}$ matrix (up to $\sim 3 \cdot 10^{22}$ $cm^{-3}$ [33,34]), and (ii) the presence in the CM M/NC/M structures the high-resistance layer of amorphous/nanocrystalline $LiNbO_3$ with $\varepsilon_d$ reaching $\sim (10^2–10^4)$ [44] which is formed near the bottom electrode during the NC growth under a large excess of oxygen (see Fig. 2b).

Fig. 6 illustrates the qualitative RS model under the above stated conditions. In a pristine state the dispersed atoms are uniformly distributed in the isolating matrix (Fig. 6a). At the first stage of switching, there is a "soft" electroforming, i.e. after applying a voltage and passing a current, the nucleation of dispersed atoms around chains of granules producing percolation paths can occur, and as a consequence the formation of metalized granular chains or MGCs. It is obvious that the manifestation of the nucleation effects is due to the strong oversaturation of the considered system by dispersed atoms and the presence of metallic nuclei (nanoparticles) in it. Previously, similar effects were observed, for example, in $SiO_2$-based memristive structures with dispersed Pt or W atoms [55, 56].

The appearing MGCs in Fig. 6 are shown isolated for simplicity, i.e. as an array of 1D nanowires. In fact, it is well known that in the case of hopping (tunneling) electron transport, the conductivity of disordered system is percolation in nature and is determined by the percolation network with a characteristic size equal to the correlation radius $L$ of the percolation cluster [57]. In the case of NC, the correlation radius is noticeably greater than the granule size, $L \sim 10$ nm (see Sec. A in [45]). Therefore, when the NC layer thickness $d_{nc} \gg L$, numerous conducting connections can be formed between the MGCs. However, these links should not inhibit the multifilamentary character of RS (see the next Sec. E).

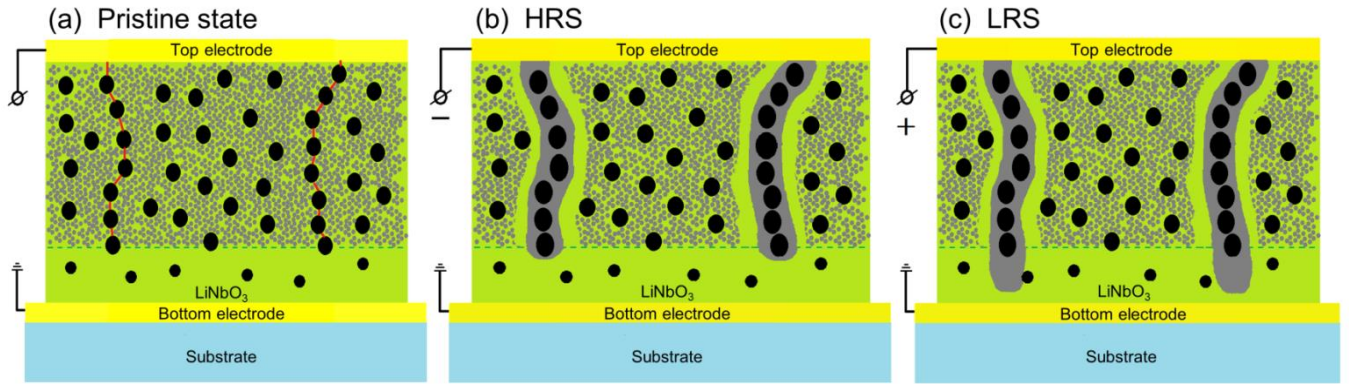

**Fig. 6.** The qualitative model of RS in the M/NC/M structures. (a) The M/NC/M structure in a pristine state after synthesis. Green color shows amorphous/nanocrystalline $LiNbO_{3-y}$ matrix, containing metal CoFe nanogranules (black ovals) and nonequilibrium phase of Co and Fe atoms with a concentration reaching $n_i \sim 10^{22}$ cm$^{-3}$ (gray circles). The red lines show the percolation paths that determine the current after applying a voltage to the structure. The green dashed line separates the high resistance layer near the bottom electrode of the structure from predominantly stoichiometric $LiNbO_3$, in which there is no nonequilibrium atomic metal phase. (b) The M/NC/M structure in HRS after soft electroforming (see text) and applying negative potential to the top electrode. The gray areas surrounding the chains of granules are metallic condensate, which occurs due to the nucleation processes of Co and Fe atoms and oxygen vacancies when the current flows through the structure. (c) The M/NC/M structure is in the LRS after applying positive potential above a certain value to the top electrode.

At the second stage, regular reversible switching is possible, but only after the soft electroforming process has taken place. At a small content of dispersed metal atoms ($n_i \sim 10^{20}$ cm$^{-3}$), when the nucleation is insignificant, there are no repeated regular RS [43]. Otherwise, when a sufficiently large negative voltage is applied to the top electrode, the structure switches to the HRS due to the movement of oxygen vacancies (cations) to the top electrode by means of MGCs and increasing the effective gap $l_g$ between the MGCs and the bottom electrode (Fig. 6b). Obviously, the capacitance of the structure in this situation must fall, since in a rough approximation the value of $C \propto 1/l_g$. The reverse situation arises when a sufficiently large positive voltage is applied to the structure (Fig. 6c). In this case, the migration of vacancies (cations) from MGCs to the bottom electrode during switching the structure to the LRS leads to a reduction of the gap, but not to its complete "collapse". This is probably due both to the relatively thick oxide layer and to the considerable resistance of MGCs, which causes the redistribution of voltage drop between $R_s$ and $R_c$ during switching the structure to the LRS (see inset to Fig. 3) and, as a consequence, the absence of short circuits between the MGCs and the bottom electrode.

Simple estimates confirm the reasonableness of the proposed RS model. Assuming $l_g \sim 1$ nm in LRS, $\varepsilon_d \sim 10^3$, lateral MGCs size $S_c^{1/2}$ of ~10 nm, and their number $N_c \sim 10^7$ on the area $S = 10^{-3}$ cm$^2$ (corresponds to the distance between MGCs ~100 nm and fill factor $K_c = N_c S_c/S \sim 10^{-2}$), we obtain the structure capacity $C \sim 9000$ pF, which is consistent with the results of the experiment: $C \approx 9000$ pF at $f \approx 400$ Hz (Fig. 4a).

Note that the proposed model is qualitative. For the development of the microscopic model of RS one needs to take into account several phenomena such as ionic, electronic, phonon transport on a nanoscale as well as effects of atomic nucleation and percolation. However, such a complex study lies beyond the scope of this work.

The proof of the proposed model could be obtained from direct TEM/STEM observations of filaments that form at the BE during RS to LRS. However, in MOM structures with RS based on the valence change mechanism with the small surface filling of filaments ($K_s \sim 1$ % or less), their direct observation is an extremely difficult task. At present, there are few works on their TEM/STEM observation, obtained mainly on thin structures with defects of the TE that occur due to the release of O$_2$ at RS [12, 13, 31, 47]. In this situation, to obtain further insight into peculiarities of the RS that confirm our model, we performed additional experiments described below.

### E. Breakdown and resistive switching of thin two-layer M/NC/LiNbO$_3$/M structures

From the model presented above, it follows that the main reason for the multifilamentary RS regime is the percolation character of the conductivity of NC, which together with the TE forms an electrical nanowires type contacts to the LiNbO$_3$ interlayer at the BE that determines the RS of M/NC/M structure. When the NC film thickness $d_{nc}$ decreases, it can be expected that the transverse conductivity of the film will begin to be determined by isolated chains of granules with an optimal low resistance (based on the analogy with the conductivity of amorphous semiconductor films [57-60]). It is also obvious that the number of such chains can begin to increase sharply with increasing $x$ under conditions when the correlation radius $L \leq d_{nc}$ (while the structure area $S \gg L^2$). Recall [33] that $L \sim (a_g+b_t) \cdot (b_t/\lambda)^v$, where $b_t$ is the effective tunneling intergranular distance ($b_t \propto a_g(x_p-x)/x_p$ at $(x_p-x) \ll x_p$), $\lambda$ is the depth of subbarrier penetration of the electron wavefunction, and $v = 0.88$ is the critical index of the percolation theory [57]. In other words, the correlation radius for hopping conductivity decreases with increasing $x$, tending to zero at the percolation threshold. (In fact, the radius tends to a certain finite value due to the finite width of the percolation threshold, where the conductivity of the medium is determined by both the "dielectric" and "metal" components [61]). Therefore, if the model described above is valid, we can expect the manifestation of transition to multifilamentary RS regime in M/NC/M structures with a thin NC layer at increasing $x$.

We used in experiments M/NC/LiNbO$_3$/M structures with $x$ = 6-20 at.%, $d_{nc} \approx$ 10 nm and the LiNbO$_3$ layer thickness of $d_{ox} \approx$ 40 nm, which is usually used to fabricate M/LiNbO$_3$/M memristors [62]. As we have already noted, one of the natural manifestations of the multifilament RS mechanism (see Sec. A above) is associated with the strong increase in switching currents, in particular, the limiting ones preceding the electrical breakdown of the structure. To find the breakdown current $I_b$, the I-V characteristics were measured by scanning the current before the irreversible transition of the M/NC/LiNbO$_3$/M structure to a conducting state with the resistance of about 10 Ω (see the bottom inset to Fig. 7). The measurements were performed at different current sweep rates determined by the value of the current change step (0.1 µA and 1.0 µA) with the same step duration of 20 ms.

Fig. 7 shows the dependence of the $I_b$ value versus $x$ in the NC layer for M/NC/LiNbO$_3$/M structure. The presented data clearly indicate the existence of two breakdown mechanisms, one of which is realized when the metal phase content is less than 13.5 at.%. The limit current of irreversible RS is in this case $I_b \approx$ 20-30 µA. However, at high values of $x \geq$ 13.5 at.%, a sharp (almost 4 orders of magnitude) increase in $I_b$ up to ~100 mA is observed. The found effect of the strong increase in the limiting current at RS with the increasing of $x$, we naturally associate with the manifestation of multifilament RS mechanism. Note that at measuring I-V curves under current-limiting conditions $I_c < I_b$, reversible RS is observed in the region of relatively low voltages: $|U| \leq$

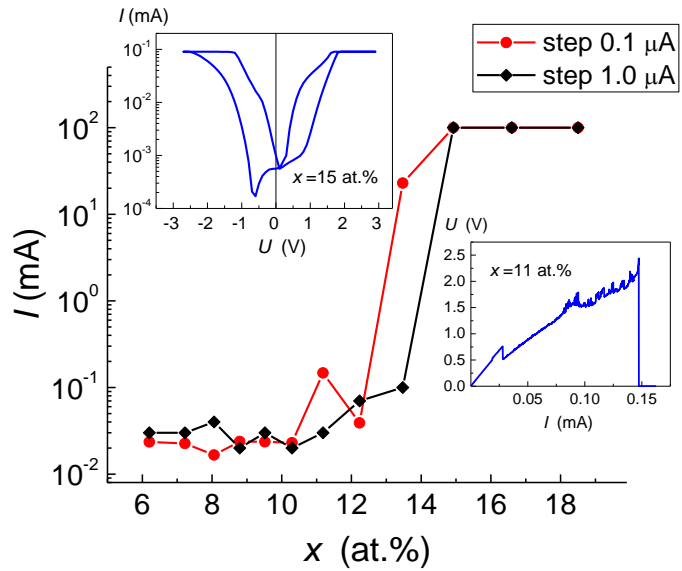

**Fig. 7.** The dependence of the breakdown current $I_b$ versus $x$ for M/NC/LiNbO$_3$/M structure measured at different current sweep rates determined by the value of the current change step (0.1 µA and 1.0 µA). The bottom and top insets show I-V curves measured by scanning the current and voltage, respectively.

2V at $I_c = 0.1$ mA, $x \approx 15$ at.% (see top inset in Fig. 7). The observed minimum on the I-V curve at negative $U$ is probably due to the manifestation of a nano-battery effect or electric polarization of the LiNbO$_3$ layer (see [63] and Fig. 2 in this work).

### F. Magnetization before and after RS of MM M/NC/M structures

Let's now consider the behavior of M/NC/M structure magnetization $M$ at its transitions from one resistive state to another. According to the picture described above (see Fig. 6), the maximum concentration $n_i$ of dispersed metal atoms (magnetic ions) is expected in the pristine state of the structure. At the same time, the maximum magnetic moment per ion is reached: $m_i = 5.9$, 5.4, and 4.8·$\mu_B$ for $Fe^{3+}$, $Fe^{2+}$ and $Co^{2+}$, respectively [64] (here $\mu_B$ is a Bohr magneton). Meanwhile, when Co and Fe atoms are nucleated into metal granules, their magnetic moment must fall: $m_m = 1.72$, 2.22, and 2.33·$\mu_B$ for Co, Fe, and CoFe, respectively [64, 65]. Therefore, at the M/NC/M structure transition from the pristine state (Fig. 6a) to the state with stable RS after soft electroforming (Fig. 6b, c), the maximum change in the magnetic moment $\Delta J$ is expected. Neglecting the effects of spin loss (gain) due to electro-thermally-accelerated oxidation (reduction) processes during electroforming [50], for the change of $\Delta J$ we have: $\Delta J = (M_{PS} - M_{RS})S_{nc}d_{nc} \approx n_i V_c (m_i - m_m) = n_i (n_s N_c S_c d_{nc})(m_i - m_m)$, where $M_{PS}$, $M_{RS}$ are magnetization before and after RS processes, respectively, $n_s = K_s \cdot S_{nc}/S$ is the number of M/NC/M structures on the NC film with an area of $S_{nc}$, and $V_c$ is the volume of MGC$_s$. Taking into account the presence of a ferromagnetic component in the magnetization $M_{FM}$ of the film, in addition to the paramagnetic component $M_{PM}$, we obtain a relative change in the magnetization after electroforming:

$$(M_{PS} - M_{RS})/M_{PS} \approx \left(\frac{N_c S_c}{S}\right)\left(\frac{m_i - m_m}{m_i}\right)\left(\frac{K_s}{1 + M_{FM}/M_{PM}}\right). \quad (4)$$

In our case, the change in the magnetic moment per atom is $\Delta m = (m_i - m_m)/m_i \approx 0.6$. It follows from (4) that in this case, the change in magnetization coincides in order of magnitude with the MGC fill coefficient, $K_c = N_c S_c/S$, for NC films with $K_s \approx 1$, in which the paramagnetic component is noticeably higher than the ferromagnetic one.

For magnetic measurements, we used MM M/NC/M structures with Cu contacts (see Sec. II), which were placed in an amount of about 50 pieces on the 0.2 cm$^2$ substrate covered with the NC layer. The freshly prepared samples were measured using SQUID magnetometer at room and helium temperatures ($T \leq 10$ K). Then the samples were taken out of the magnetometer and each of the sample structures was subjected to 5-6 cyclic RS, which ended when the structures returned to the LRS. Then the magnetization was measured again in these samples. By measuring samples with $x \approx$ 18-19 at.% above percolation threshold in which there are no RS, we estimated that in our case the systematic error of measuring the magnetic moment $J$ does not exceed $\delta J \approx \pm 10^{-4}$ emu when the sample $J$ is $\approx 2 \cdot 10^{-2}$ emu.

Figure 8a shows the magnetic field dependences of magnetization measured at $T = 300$ K and $T \leq 10$ K for MM M/NC/M structures with $x \approx 15$ at.%. From the presented data, it follows that at low temperatures there is a noticeable contribution to the $M$ value of PM component, which is 1.6 times higher than the contribution of the FM one at 300 K. Fitting difference of the field dependences of magnetization $M(H,T)$ measured at several temperatures from the low temperature range ($T \leq 10$ K) by difference of Brillouin's functions (see details in [34]), we find the concentration of dispersed ions for this sample to be $n_i \approx 6.5 \cdot 10^{21}$ cm$^{-3}$. M/NC/M structures with such content of dispersed atoms demonstrate stable RS [43] (see I-V curves in the inset to Fig. 8a).

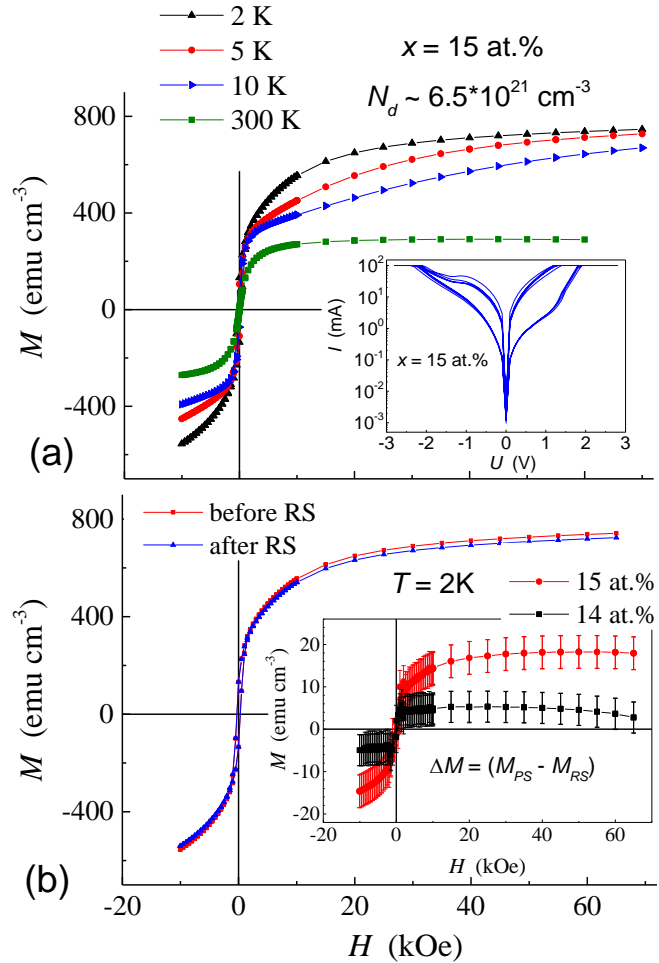

**Fig. 8**. The magnetic field dependences of magnetization $M(H)$ for MM M/NC/M structures with $x \approx$ 15 at.% measured at different temperatures (a), and at $T = 2$ K before and after cyclic RS (b). The top inset shows 6 cyclic I-V curves; the bottom one demonstrates dependencies of magnetization reduction $\Delta M(H) = (M_{PS} - M_{RS})$ after RS.

Fig. 8b shows the dependencies $M(H)$ obtained at $T = 2$ K for the sample with $x \approx 15$ at.% before and after RS. It can be seen that there is a slight reduction in the $\Delta M = (M_{PS} - M_{RS})$ after RS. Examples of $\Delta M(H)$ dependencies are shown in the inset to Fig. 8b for samples with $x \approx 14$ and 15 at.%. In both cases, the value of $\Delta M \approx$ (5-15) emu/cm$^3$, although small, clearly goes beyond the systematic error of measurements $\delta M \approx \pm 4$ emu/cm$^3$. According to data presented in Fig. 8b, the relative decrease in magnetization $\Delta M/M_{PS} \sim 10^{-2}$, which corresponds according to (4) to the MGC fill factor $K_c \sim 10^{-1}$, taking into account that in our case $K_s \approx 0.25$, $M_{PM}/M_{FM} \approx 1.6$, and $\Delta m/m_i \approx 0.6$.

In Sec. D, an estimation of $K_c \sim 10^{-2}$ is obtained based on capacitance measurements, assuming that the permittivity of amorphous LiNbO$_3$ $\varepsilon_d \sim 10^3$. However, this $\varepsilon_d$ value corresponds to LiNbO$_3$ films that were heated to the $T \approx 100$-$200$ $^o$C [44]. At room temperature for amorphous LiNbO$_3$ the value of $\varepsilon_d \approx 50$-$90$ [44], which corresponds to an order of magnitude lower values of $K_c \sim 10^{-1}$. In addition, it is necessary to keep in mind possible discrepancies between the data of magnetic and capacitive measurements related to the effects of thermally-accelerated oxidation (reduction) [50], which were not taken into account in the derivation of (4).

Note also that after soft electroforming, in our case, a ferromagnet/insulator/non-magnetic metal junction is formed at the BE, which does not exhibit spin-dependent magnetoresistance [66], in contrast to the ferromagnet/insulator/ferromagnet junctions that occur, for example, under RS of M/NiO/M structures [31]. Therefore, at this stage, we have limited ourselves to the study only of the effect of RS on magnetization.

### G. Temperature dependencies of resistance for CM M/NC/M structures

To obtain further insight into peculiarities of the RS, confirming our model, the temperature dependencies of resistance $R(T)$ were investigated for structure in both the HRS and LRS. Switching to the LRS was carried out at the temperature of 250 K (Fig. 9a) in order to avoid the influence of possible conductance relaxations on the dependence of $R(T)$. Note, that the RS were not observed below 200 K in the whole voltage range from −15 to +15 V. Under these conditions the I–V curves were symmetrical and nonlinear (Fig. 9b).

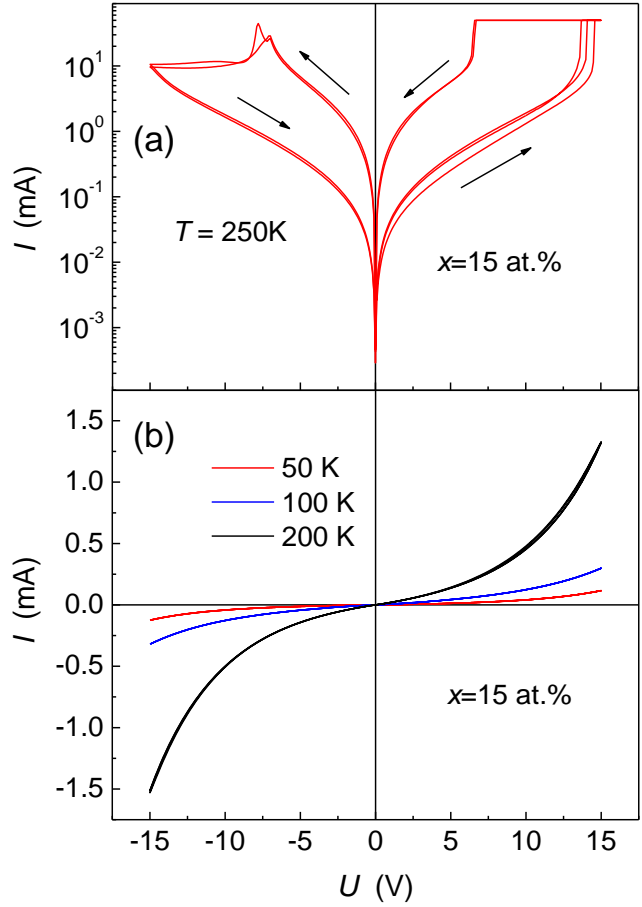

**Fig. 9.** Typical I–V curves measured for the M/(CoFeB)$_x$(LiNbO$_3$)$_{100-x}$/M structure with $x \approx 15$ at. % at (a) $T = 250$ K and (b) $T = 50, 100$ and $200$ K.

Fig. 10 shows the $R(T)$ dependencies in the HRS and after switching the structure to LRS, as a result of which its resistance decreases by 4 times. In both cases, the activation dependences are observed, moreover, in a wide temperature range ($T = 15–120$ K) they are approximated with high accuracy (better than 1%) by the function $\ln R$ on $(1/T)^{1/2}$. For comparison, the inset to Fig. 10 also shows the dependence of $\ln R$ vs $1/T$, which is not linear in the studied temperature range. In other words, the conductance follows the law of $\ln R \propto (T_0/T)^{1/2}$, often observed in systems with hopping conductivity [33, 57], and with almost identical values of the parameter $T_0 \approx 1440$ K (in HRS) and 1410 K (in LRS). This clearly indicates that in our case RS of the structure to lower resistance state is accompanied by a decrease in the $l_g$ value and absence of metal short circuits. The large value of $T_0$ is obviously related with the large energy depth of traps in the oxide and the strong localization of electrons on them. Indeed, in the case of doped semiconductors (or highly defective dielectrics), the law "1/2" in the $R(T)$ dependence is associated with the formation of a Coulomb gap near the Fermi level. The parameter $T_0$ is determined by the localization radius $a_B$ of the electron at the defect [57]: $k_B T_0 \sim e^2/\varepsilon_d a_B$, where $k_B$ is Boltzmann's constant, $e$ is the charge of the electron. In the case of the strong localization (of about the lattice constant), the permittivity at optical frequencies should be used to estimate $a_B$, assuming $\varepsilon_d \approx \varepsilon_{opt}$. In LiNbO$_3$ the value of $\varepsilon_{opt}$ is about 10 [67]. Using the expression for $T_0$ and the value found in the experiment of $T_0 \sim 1400$ K, we obtain $a_B \sim 1$ nm, which is a reasonable result (hexagonal lattice parameters for LiNbO$_3$: $a = 0.5148$ nm and $c = 1.3863$ nm [67]).

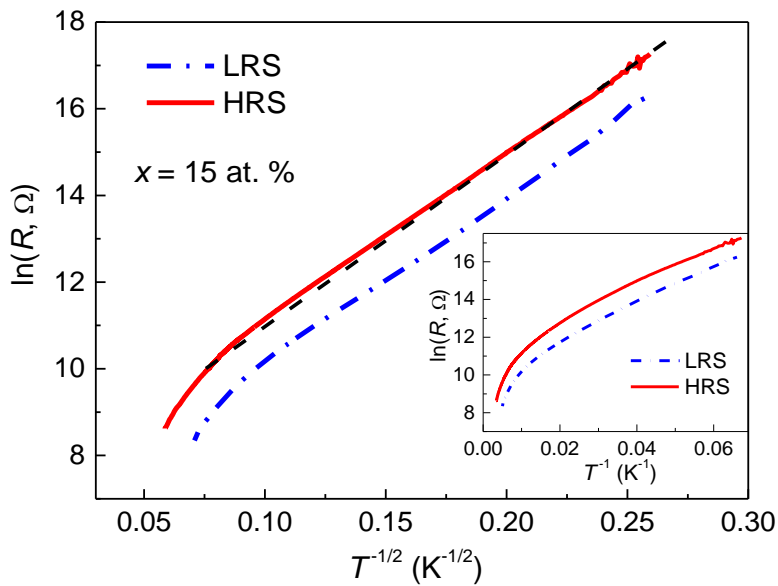

**Fig. 10.** The temperature dependencies of resistance $R(T)$ for the M/(CoFeB)$_x$(LiNbO$_3$)$_{100-x}$/M structure in the two different resistive states in coordinates of $\ln R$ vs $(1/T)^{1/2}$. The inset shows $\ln R$ as a function of $1/T$.

## IV. CONCLUSION

Thus, the experimental results presented above and their analysis show that the originality of the memristive properties of the M/NC/M samples based on the (Co$_{40}$Fe$_{40}$B$_{20}$)$_x$(LiNbO$_3$)$_{100-x}$ nanocomposite is determined both by their structural features associated with the formation of high-resistance and highly polarizable (high-$\kappa$) oxide interface layer and by the multifilamentary nature of the resistive switching. The latter is due to the presence of a large number of granules in the oxide matrix as well as dispersed metal atoms in it, whose nucleation causes formation of many channels of resistive switching. As a result of synergetic combination of these factors, not only a significant change in the structure resistance is observed, but also in its capacity, which reaches 8 times at the ratio of $R_{OFF}/R_{ON} \sim 40$.

The proposed qualitative model of multifilamentary character of RS allows explaining high level of endurance (more than $10^6$) and plasticity of our samples (the possibility of a smooth change of the resistance in the ($R_{OFF} - R_{ON}$) window, more than 300 resistive states), which, in particular, allow emulating the unique properties of biological synapses using them. The obtained results pave the way for creating hardware neural networks based on nanocomposite memristors with controllable characteristics.

Finally, we note that the proposed multifilamentary RS model is confirmed by investigations of breakdown and RS of two-layer nanoscale M/NC/LiNbO$_3$/M structures, studies of the magnetization of M/NC/M structures in the pristine state and after RS, as well as the study of resistance temperature dependences of these structures in various resistive states. However, additional experiments are needed to study the RS mechanism of nanocomposite memristive structures aimed, in particular, at visualizing resistive switching channels using methods of high-resolution electron microscopy.

**Acknowledgments**

This work was supported by the Russian Science Foundation (project No. 16-19-10233). Measurements were carried out with the equipment of the Resource Centers (NRC "Kurchatov Institute").

**Figure captions**

Fig.1. (a) The dependence of conductance $G(x) = I/U$ of the CM M/NC/M structures ($d \approx 1.5$ µm) on the metal content $x$, measured at DC at $U = 0.3$ V. The inset shows the I-V curves for the CM M/NC/M structure with optimal content of $x_{opt} \approx 15$ at. %. (b) The I-V characteristics of the M/NC/M structures with the NC thickness of $d \approx 2.5$ µm and metal content of $x \approx 8$ at. % obtained by 30 cyclic measurements. The inset demonstrates the endurance to RS of this structure, i.e. the dependence of resistance after set and reset pulses on the switching cycle number. The arrows on the insets show the voltage scanning direction.

Fig.2. The cross-section image, results of EDX analysis and size distribution of granules of CM M/NC/M structure with optimal value $x_{opt} \approx 15$ at. %. (a) Bright field STEM image of the sample. (b) Enlarged HAADF STEM image of the area near bottom electrode. The white line demonstrates the area of EDX element analysis, shown in (c). (d), (e) HAADF STEM images of the areas marked in (b) with red and yellow squares, respectively. (f), (g) Histogram of the transverse-size distribution of granules and Gaussian approximation of this dependence (solid curve) for the images shown in (d) and (e), respectively. (h) HRTEM image of the interface near the bottom electrode. (i) The 2D FFT spectrum corresponding to the CoFe granule marked in (h) by the red arrow. (k) The image of the granule crystal lattice in [001] zone axis.

Fig.3. The dependences of the imaginary part of the impedance on the real one for the CM M/NC/M structure with $x \approx 15$ at. % in the HRS and LRS (the top inset). The bottom inset shows the equivalent circuit of the CM M/NC/M structure (explanations are given in the text).

Fig.4. The frequency dependences of capacitance (a) and resistance (b) in the HRS and LRS for the M/(CoFeB)$_x$(LiNbO$_3$)$_{100-x}$/M structure with $x \approx 15$ at. %. The inset shows the frequency dependences of loss tangent in the HRS and LRS for this structure.

Fig.5. Dependence of the relative change in resistance and capacity of the CM M/NC/M structure during its RS between the HRS and LRS on the content of metallic phase at a frequency of 1 kHz.

Fig.6. The qualitative model of RS in the M/NC/M structures. (a) The M/NC/M structure in a pristine state after synthesis. Green color shows amorphous/nanocrystalline LiNbO$_{3-y}$ matrix, containing metal CoFe nanogranules (black ovals) and nonequilibrium phase of Co and Fe atoms with a concentration reaching $n_i \sim 10^{22}$ cm$^{-3}$ (gray circles). The red lines show the percolation paths that determine the current after applying a voltage to the structure. The green dashed line separates the high resistance layer near the bottom electrode of the structure from predominantly stoichiometric LiNbO$_3$, in which there is no nonequilibrium atomic metal phase. (b) The M/NC/M structure in HRS after soft electroforming (see text) and applying negative potential to the top electrode. The gray areas surrounding the chains of granules are metallic condensate, which occurs due to the nucleation processes of Co and Fe atoms and oxygen vacancies when the current flows through the structure. (c) The M/NC/M structure is in the LRS after applying positive potential above a certain value to the top electrode.

Fig.7. The dependence of the breakdown current $I_b$ versus $x$ for M/NC/LiNbO$_3$/M structure measured at different current sweep rates determined by the value of the current change step (0.1 µA and 1.0 µA). The bottom and top insets show I-V curves measured by scanning the current and voltage, respectively.

Fig.8. The magnetic field dependences of magnetization $M(H)$ for MM M/NC/M structures with $x \approx 15$ at.% measured at different temperatures (a), and at $T = 2$ K before and after cyclic RS (b). The top inset shows 6 cyclic I-V curves; the bottom one demonstrates dependencies of magnetization reduction $\Delta M(H) = (M_{PS} - M_{RS})$ after RS.

Fig.9. Typical I–V curves measured for the M/(CoFeB)$_x$(LiNbO$_3$)$_{100-x}$/M structure with $x \approx 15$ at. % at (a) $T = 250$ K and (b) $T = 50, 100$ and $200$ K.

Fig.10. The temperature dependencies of resistance $R(T)$ for the M/(CoFeB)$_x$(LiNbO$_3$)$_{100-x}$/M structure in the two different resistive states in coordinates of $\ln R$ vs $(1/T)^{1/2}$. The inset shows $\ln R$ as a function of $1/T$.